\documentstyle[12pt,a4wide]{article}

\newcommand{\news}{\setcounter{equation}{0}}
\newcommand{\be}{\begin{equation}}
\newcommand{\ee}{\end{equation}}
\newcommand{\bea}{\begin{eqnarray}}
\newcommand{\eea}{\end{eqnarray}}
\newcommand{\bean}{\begin{eqnarray*}}
\newcommand{\eean}{\end{eqnarray*}}
\font\upright=cmu10 scaled\magstep1
\font\sans=cmss12
\newcommand{\ssf}{\sans}
\newcommand{\stroke}{\vrule height8pt width0.4pt depth-0.1pt}
\newcommand{\Z}{\hbox{\upright\rlap{\ssf Z}\kern 2.7pt {\ssf Z}}}

\newcommand{\C}{{\rlap{\rlap{C}\kern 3.8pt\stroke}\phantom{C}}}
\newcommand{\R}{\hbox{\upright\rlap{I}\kern 1.7pt R}}
\newcommand{\CP}{\C{\upright\rlap{I}\kern 1.7pt P}}
\newcommand{\half}{\frac{1}{2}}
\newcommand{\mt}{\rlap{\ssf T}\kern 3.0pt{\ssf T}}
\newcommand{\spc}{spectral curve }

\newcommand{\identity}{{\upright\rlap{1}\kern 2.0pt 1}}

\newcommand{\g}{\hskip 1cm}
\newcommand{\st}{\widetilde s}
\newcommand{\rt}{\sqrt{2}}
\input epsf

\begin{document}
\pagestyle{plain}
\date{August 1996}

\title{
\begin{flushright}
{\normalsize UKC/IMS/96-52}\\ 
{\normalsize \ } \\
\end{flushright}
\vskip 20pt
{\huge \bf Cyclic Monopoles} \vskip 20pt}
\author{Paul M. Sutcliffe
\thanks{This work was supported in part by the 
Nuffield  Foundation}\\[20pt]
{\normalsize \sl Institute of Mathematics} \\[5pt]
{\normalsize\sl University of Kent at Canterbury} \\[5pt]
{\normalsize\sl  Canterbury CT2 7NZ, England} \\[20pt]
{\normalsize\sl email\  P.M.Sutcliffe@ukc.ac.uk }\\[10pt]}

\maketitle

\begin{abstract}

We study charge $k$ $SU(2)$ BPS monopoles which are
symmetric under the cyclic group of order $k$.
Approximate twistor data (spectral curves and Nahm data) 
is constructed using a new technique based upon a
Painlev\a'e analysis of Nahm's equation around a pole.
With this data both analytical and numerical
approximate ADHMN constructions are performed to study the
zeros of the Higgs field and the monopole energy densities.
The results describe, via the moduli space approximation,
a novel type of low energy $k$ monopole scattering.

\end{abstract}

\newpage

\section{Introduction}
\news

By studying rational maps it has been shown \cite{HMM}
that the imposition of cyclic $C_k$ symmetry upon
strongly centred charge $k$ $SU(2)$ BPS monopoles
selects out from the general $k$-monopole moduli space
${\cal M}_k$ a set of $k$ 2-dimensional totally geodesic
submanifolds. Each of these 2-dimensional submanifolds
is a surface of revolution, hence by application of a 
reflection symmetry a number of geodesics in
${\cal M}_k$ are obtained. In the moduli space approximation
\cite{M} these geodesics may be interpreted in terms of
the motion of $k$ slowly moving monopoles. It is therefore of
considerable interest to know more about these one-parameter
families of monopoles. The rational map approach is limited
for this purpose, and the only additional information it
supplies concerns the asymptotic in and out states for the
scattering process.

Other twistor approaches yield more information about
the monopole, but are more difficult to apply than the
rational map method. In \cite{HMM} only the form of the corresponding
spectral curves was found. In this paper we do a little better, by
constructing approximate spectral curves, which of course have this
required form. Moreover, we construct approximate Nahm data and
implement an ADHMN construction to compute the Higgs field and 
energy density of these monopoles.

 The technique used is to make use of the integrability of Nahm's
equation, not to attempt an explicit solution, but rather to
study the nature of a solution around a given pole, with the aid
of a little Painlev\a'e analysis. The method applies equally well to
all values of $k$, but only the cases $k=3$ and
$k=4$ will be discussed in detail.

\section{Monopoles and ADHMN}
\news

$SU(2)$ BPS monopoles are topological soliton solutions of
 a Yang-Mills-Higgs gauge theory with no Higgs self-coupling.
Boundary conditions imply that the Higgs field at infinity
defines a map between two-spheres. This map has an integer valued
winding number $k$, which we identity as the magnetic charge or
number of monopoles. In this paper we shall be concerned with
strongly centred monopoles, which basically means that we fix
the centre of mass of the monopole configuration to be the origin,
and set a total phase to be unity. For more precise details
of strong centring and other background information on monopoles
we refer the reader to \cite{HMM,AH}. 

As mentioned in the introduction, requiring invariance of
a $k$-monopole under cyclic
$C_k$ symmetry, and an additional reflection symmetry, leads to
a number of geodesics $\Sigma_k^l$ (we follow the notation of
\cite{HMM}) in the $k$-monopole moduli space ${\cal M}_k$.
Essentially there are $(2k+3+(-1)^k)/4$ different types of
these cyclic geodesics, corresponding to $l=0,1,..k/2$ if $k$ is even
and $l=0,1,..(k-1)/2$ if $k$ is odd. Physically, for $l\ne 0$,
the associated
monopole scatterings are distinguished by having the out state
(or in state by time reversal) consisting of two clusters of
monopoles with charges $k-l$ and $l$. This explains why we do not
allow $l>k/2$, since this is basically the same scattering event as one
of the geodesics with $l<k/2$. 
If $l=0$ then the monopoles remain in a plane and scatter
instantaneously through the axisymmetric $k$-monopole and emerge
with a $\pi/k$ rotation. This kind of $\pi/k$ scattering is
essentially a two-dimensional process and has been extensively
studied in planar systems \cite{KPZ}. In this paper we shall be
concerned with the more exotic scatterings with $l\ne 0$.
It should also be pointed out that
the case $k=2$ is special, in that the geodesics $\Sigma_2^0$ and
$\Sigma_2^1$ are isomorphic, so that there is only one type
of scattering; the famous right-angle scattering found by
 Atiyah and Hitchin \cite{AH}.

In this paper we shall only deal with the cases $k=3$ and $k=4$.
The method applies equally well to all values of $k$, but all the
important features are captured by these two examples.
From the above we see that for $k=3$ there is only one interesting
geodesic $\Sigma_3^1$. It describes three individual monopoles which
scatter and emerge as a single monopole and a 2-monopole cluster.
It also contains the tetrahedral 3-monopole \cite{HMM,HSa} as
an instantaneous configuration. For $k=4$ there are two interesting
geodesics $\Sigma_4^2$ and $\Sigma_4^1$. The first describes
a scattering in which the monopoles emerge as two 2-monopole 
clusters, and also includes the cubic 4-monopole \cite{HMM,HSa}.
The second describes a scattering which results in the 
formation of a single monopole and a 3-monopole cluster; no monopoles
with the symmetries of a Platonic solid are contained in this geodesic.

There are several twistor techniques which are applicable to monopoles, 
but here our main tool will be the ADHMN \cite{Na,Hb} 
construction.
This is an equivalence between $k$-monopoles and Nahm data
$(T_1,T_2,T_3)$, which are three $k\times k$ matrices which depend
on a real parameter $s\in[0,2]$ and satisfy the following;\\

\newcounter{con}
\setcounter{con}{1}
(\roman{con})  Nahm's equation
\be
\frac{dT_i}{ds}=\half\epsilon_{ijk}[T_j,T_k] \nonumber
\ee\\

\addtocounter{con}{1}
(\roman{con}) $T_i(s)$ is regular for $s\in(0,2)$ and has simple
poles at $s=0$ and $s=2$,\\

\addtocounter{con}{1}
(\roman{con}) the matrix residues of $(T_1,T_2,T_3)$ at each
pole form the irreducible $k$-dimensional representation of SU(2),\\

\addtocounter{con}{1}
(\roman{con}) $T_i(s)=-T_i^\dagger(s)$,\\

\addtocounter{con}{1}
(\roman{con}) $T_i(s)=T_i^t(2-s)$.\\

\setcounter{con}{1}
Equation (\roman{con}) is equivalent to a Lax pair and hence there
is an associated algebraic curve, which is in fact the spectral curve
\cite{Hb}.
Explicitly, the \spc may be read off from the Nahm data as the
equation
\be
\mbox{det}(\eta+(T_1+iT_2)-2iT_3\zeta+(T_1-iT_2)\zeta^2)=0.
\label{curve}
\ee
It is useful to construct the spectral curve since it not
only gives a convenient representation of the monopole, but
also furnishes the constants for Nahm's equation.

The procedure by which the Higgs field (and gauge potential)
can be reconstructed from the Nahm data is outlined in section 5,
where our analytical approximation to this procedure is also
discussed.

In the next section we give the general form for
 cyclic $k$-monopole Nahm data and introduce the relevant
equations for the case $k=3$.

\section{Cyclic Nahm data}
\news

As explained in \cite{Sa} the form of the Nahm data for
$C_k$ symmetric $k$-monopoles may be obtained
as a linear sum of generators of the affine Lie algebra 
$A_{k-1}^{(1)}$. Explicitly, consider the Lie algebra
$A_{k-1}^{(1)}$, with $H_i$, $i=0,..,k-1$ the generators of the
extended Cartan subalgebra and $E_{\pm i}$ the generators 
corresponding to the simple roots $\alpha_i$, $i=1,..,k-1$,
plus the lowest root $\alpha_0=-\sum_{j=1}^{k-1}\alpha_j$.
In the Chevalley basis these satisfy 
\bea
&[H_i,E_{\pm j}]&=\pm C_{ij}E_{\pm j} \\
&[E_i,E_{-j}]&=\delta_{ij}H_j
\eea
where $C_{ij}$ are the elements of the $k\times k$
extended Cartan matrix given by
\be
C_{ij}=\frac{2(\alpha_i,\alpha_j)}{(\alpha_i,\alpha_i)},
\hskip 1cm i,j=0,..,k-1.
\ee
Cyclic Nahm data may be expressed in terms of
a linear sum of generators as
\be
T_1-iT_2=\sum_{j=0}^{k-1}Q_jE_{+j}, \hskip 1cm
T_3=i\sum_{j=0}^{k-1}P_jH_j
\label{toda}
\ee
with real function coefficients $P_j,Q_j$.

For the case $k=3$ a change of notation yields
the Nahm data for $C_3$ symmetric 3-monopoles to have the form
$$
T_1=\frac{1}{2}\left[\begin{array}{ccc}
0&-f_3&f_2\\
f_3&0&-f_1\\
-f_2&f_1&0\end{array}\right]; \
T_2=-\frac{i}{2}\left[\begin{array}{ccc}
0&f_3&f_2\\
f_3&0&f_1\\
f_2&f_1&0\end{array}\right]; \
$$
\be
T_3=i\left[\begin{array}{ccc}
f_4&0&0\\
0&f_5&0\\
0&0&-(f_4+f_5)\end{array}\right]
\ee
with corresponding 
 spectral curve 
\be
\eta^3+\alpha\eta\zeta^2+\gamma\zeta^3
+i\beta(\zeta^6+1)=0.\ee
The constants $\alpha$, $\beta$, $\gamma$ can all
be taken to be real, by imposition of a reflection 
symmetry, and are given in terms of the $f_i$'s as
\bea
&\alpha=&f_1^2+f_2^2+f_3^3-4(f_4^2+f_5^2+f_4f_5)
\nonumber \\
&\beta=&-f_1f_2f_3 \nonumber \\
&\gamma=&2f_1^2f_4+2f_2^2f_5-2(f_3^2+4f_4f_5)(f_4+f_5).
\eea
For this data Nahm's equation becomes
\bea
&\dot f_1=&(2f_5+f_4)f_1 \nonumber \\
&\dot f_2=&-(2f_4+f_5)f_2  \nonumber \\
&\dot f_3=&(f_4-f_5)f_3  \nonumber \\
&\dot f_4=&(f_3^2-f_2^2)/2 \nonumber \\
&\dot f_5=&(f_1^2-f_3^2)/2.
\label{feq}
\eea
The first calculation we require is to see how the
tetrahedral 3-monopole sits inside this Nahm data.
The spectral curve and Nahm data of the tetrahedral 
3-monopole have been computed \cite{HMM} for the monopole
in a different orientation and it is possible to 
derive the answer we require by rotation of this known
data. However, this is a non-trivial exercise which in fact
requires more work than simply solving the equations again.
Thus in the remainder of this 
 section we compute the Nahm data and
spectral curve of the tetrahedral 3-monopole
in the orientation in which it has $C_3$ symmetry
around the $x_3$ axis.

For tetrahedral symmetry we must have that $\alpha=0$,
which is achieved by the following choice
\be
f_1^2=2f_4^2, \g  f_2^2=2f_5^2, \g f_3^2=2(f_4+f_5)^2.
\ee
The remaining constants then simplify to 
\be
\gamma=-20f_4f_5(f_4+f_5), \g \mbox{and} \g 
\beta=-\frac{\gamma}{5\sqrt{2}}.
\ee
Using the above constants we can solve for $f_5$ in terms
of $f_4$ as
\be
2f_5=-f_4\pm\sqrt{f_4^2-\gamma/(5f_4)}
\ee
and substituting this into the equation (\ref{feq})
for $f_4$ yields
\be
\dot f_4^2=f_4(f_4^3-\gamma/5).
\ee
Introducing the variable $2w=-1/f_4$ this equation
becomes the standard form elliptic equation
\be
4\dot w^2=cw^3+1, \g c=8\gamma/5.
\ee
Set 
\be w=\wp(t)/(4\kappa) \ee
with $t=\kappa s+B$ then, choosing
$4\kappa=c^{1/3}$, $\wp$ is the Weierstrass elliptic
function satisfying
\be \wp^{\prime 2}=4\wp^3+4 \ee
where prime denotes differentiation with respect to the
argument.
The real period of this elliptic function is
\be 
2\omega_1=\Gamma(1/6)\Gamma(1/3)/(2\sqrt{\pi}).
\ee
For the correct boundary conditions take the 
elliptic function to go between zeros at
$t=2\omega_1/3$ and $t=4\omega_1/3$.
This requires that $\kappa=\omega_1/3$ and $B=2\omega_1/3$.
Now 
\be
-\beta=\frac{\gamma}{5\sqrt{2}}
=\frac{c}{8\sqrt{2}}=\frac{(4\kappa)^3}{8\sqrt{2}}
=\Gamma(1/6)^3\Gamma(1/3)^3
/(\pi^{3/2}3^38\sqrt{2}).
\ee
Thus we have arrived at the spectral curve
\be
\eta^3-i(\zeta^6+i5\sqrt{2}\zeta^3+1)
\Gamma(1/6)^3\Gamma(1/3)^3
/(\pi^{3/2}3^38\sqrt{2})
=0.
\label{tetcurve}\ee
It is a simple, though tedious, calculation
 to verify\footnote{I thank Conor Houghton for checking this}
that this is indeed the spectral curve obtained by
rotating the one given in \cite{HMM}.

We have yet to examine the residue behaviour of the
functions, and this will be of vital importance
in what follows later.
For $t\sim B$
\be
\wp(t-B)\sim -2(t-B)
\ee and
\be 
f_4=-\frac{1}{2w}\sim \frac{4\kappa}{4(t-B)}=1/s.
\ee
The other relations then give 
$f_5\sim -1/s$, and hence
$$T_3\sim s^{-1}i\mbox{diag}(1,-1,0)$$ which identifies
the representation formed by the residues as the 
irreducible one.
As it will be needed later we list here the
residue behaviour of all the functions at both ends
of the interval,
\be
f_1\sim \sqrt{2}/s, \ f_2\sim\sqrt{2}/s, \ 
f_3\sim 0/s, \ f_4\sim 1/s, \ f_5\sim -1/s, \g
\mbox{as} \ s\rightarrow 0
\label{reslhs}
\ee
and defining the variable $\st=2-s$ then
\be
f_1\sim \sqrt{2}/\st, \ f_2\sim\ 0/\st, \ 
f_3\sim\sqrt{2}/\st, \ f_4\sim 1/\st, \ f_5\sim 0/\st, \g
\mbox{as} \ s\rightarrow 2.
\label{resrhs}
\ee

\section{Approximate twistor data}
\news
In principle the equations were are concerned with for
cyclic $k$-monopoles are solvable in terms of abelian integrals of
genus $(k-1)$ \cite{Sa}.
However, for $k>2$, to explicitly extract from the general
solution the one which satisfies all the required 
boundary conditions appears to be a highly non-trivial 
exercise. Furthermore, one of the main motivations for
constructing the Nahm data is so that it can be used as
input in the numerical ADHMN algorithm \cite{HSa}, to
provide a visualization of the monopole energy densities.
Even if the required explicit solution could be determined
it is by no means clear that it would be in a form suitable
for obtaining numerical values;
there are at present no numerical algorithms available to 
compute the Riemann theta function for a surface of genus greater
than one, unless it happens to be of a very special form
which allows Weierstrass reduction theory to be applied 
\cite{E}. With this in mind we
construct, in this section, approximate twistor data
(ie. Nahm data and spectral curves) for $C_3$ symmetric
3-monopoles.

One of the key points in applying the following method
is that we know explicitly the twistor
data for one member of the family ie. the tetrahedral
3-monopole. 
The one parameter family of monopoles
we are searching for is a geodesic in the monopole moduli
space and it is known \cite{N} that the transformation 
between the monopole moduli space metric and the metric on Nahm
data is an isometry. Since the Nahm data has poles it follows
from these two facts that the residues at these poles
must be constant with respect to the geodesic parameter.
The upshot is that we know the residue behaviour explicitly
for all members of the one-parameter family, it is given
by (\ref{reslhs}) and (\ref{resrhs}).

As equations (\ref{feq}) are integrable they possess
a particular solution which is a single-valued expansion
 around the $s=0$ pole ie
\be
f_i=\sum_{j=-1}^\infty a_{i,j}s^j
\ee
where the pole coefficients are given by (\ref{reslhs})
as
\be
a_{1,-1}=\sqrt{2}, \g a_{2,-1}=\sqrt{2}, \g
a_{3,-1}=0, \g a_{4,-1}=1, \g a_{5,-1}=-1.
\ee
We now need to determine the number of arbitrary
constants in the above particular solution.
This can be done using Painlev\a'e analysis \cite{ARS,Y}
as follows. We consider the two term expansion
\be
f_i=\frac{a_{i,-1}}{s}+b_is^{r-1}
\ee 
for $r$ an arbitrary integer, and linearise the equation
(\ref{feq}) to obtain
\be
\left[\begin{array}{ccccc}
r&0&0&-\sqrt{2}&-2\sqrt{2}\\
0&r&0&2\sqrt{2}&\sqrt{2}\\
0&0&r-3&0&0\\
0&\sqrt{2}&0&r-1&0\\
-\sqrt{2}&0&0&0&r-1
\end{array}
\right]
\left[\begin{array}{c}b_1\\b_2\\b_3\\b_4\\b_5\end{array}\right]
=\left[\begin{array}{c}0\\0\\0\\0\\0\end{array}\right].
\ee
The Kowalevski exponents (they are not strictly
resonances since $a_{3,-1}=0$) are determined by the
vanishing of the determinant $\Delta$ of the above
matrix. We have that
\be
\Delta=(r-3)^2(r-2)(r+1)(r+2)
\ee
so there are three arbitrary constants in our particular
solution, corresponding to the three (counted with 
multiplicity) positive roots of $\Delta$.
Furthermore, we see that one of the arbitrary constants
appears at the linear level in the $s$ expansion and
the remaining two at quadratic level.
Explicit calculation reveals that these may be taken to be
$a_{1,1},a_{1,2}$ and $a_{3,2}$. We obtain that, to
cubic order in $s$,
\bea
&f_1=&\rt/s+a_{1,1}s+a_{1,2}s^2+\frac{7\rt}{20}a_{1,1}^2s^3
\nonumber \\
&f_2=&\rt/s+a_{1,1}s-a_{1,2}s^2+\frac{7\rt}{20}a_{1,1}^2s^3
\nonumber \\
&f_3=&a_{3,2}s^2 \nonumber \\
&f_4=&1/s-\rt
a_{1,1}s+\frac{1}{\rt}a_{1,2}s^2-\frac{2}{5}a_{1,1}^2s^3
\nonumber \\
&f_5=&-1/s+\rt
a_{1,1}s+\frac{1}{\rt}a_{1,2}s^2+\frac{2}{5}a_{1,1}^2s^3.
\label{expan}
\eea
This is going to be the form of the functions in 
our approximate Nahm data. Note that we terminate the expansion
at $O(s^3)$, as this will turn out to be sufficient for our
needs, but it is a simple matter to improve the
accuracy of the approximate data by keeping higher order
terms in the above. 

At this stage we have a candidate three-parameter family
of approximate data, from which we need to select the correct
one-parameter family. The constraining conditions arise
from consideration of a symmetry of the equations (\ref{feq}).
As before, let $\st=2-s$, then if $f_i(s)$ is a solution
we can construct a second solution $g_i(s)$ as
\be
[g_1(s),g_2(s),g_3(s),g_4(s),g_5(s)]=
[f_1(\st),f_3(\st),f_2(\st),f_4(\st),-f_4(\st)-f_5(\st)].
\ee
Note that if $f_i(s)$ is a solution of the form
(\ref{expan}) with a pole at $s=0$ then $g_i(s)$ will
have a pole at $s=2$. Furthermore, the residues at
this pole will be precisely those of (\ref{resrhs})
which we require.
Thus, we take our approximate solution to be $f_i(s)$ for
$s\in[0,1]$ and $g_i(s)$ for $s\in[1,2]$. 
There is now the matching condition at $s=1$ that
$g_i(1)=f_i(1)$, for $i=1,..,5$. The first and fourth
of these equations are identities, while the second
and third are equivalent, so we are left with the
two matching conditions
\be
f_2(1)=f_3(1) \g \mbox{and} \g f_4(1)=-2f_5(1).
\ee
These give two relations between the three parameters
$a_{1,1},a_{1,2},a_{3,2}$ and thus determine the 
sought after one-parameter family.
Using the series (\ref{expan}) the matching conditions
give that
\bea
&a_{3,2}=&\rt+a_{1,1}-a_{1,2}+\frac{7\rt}{20}a_{1,1}^2
\nonumber \\
&a_{1,2}=&\frac{\rt}{3}(1-\rt a_{1,1}-\frac{2}{5}a_{1,1}^2)
\eea
so that everything is determined by the parameter $a_{1,1}$
which we now relabel as $a$. In detail the functions are
\bea
&f_1=&\rt/s+as+\frac{\rt}{3}(1-\rt a -\frac{2}{5}a^2)s^2
+\frac{7\rt}{20}a^2s^3 \nonumber\\
&f_2=&\rt/s+as-\frac{\rt}{3}(1-\rt a -\frac{2}{5}a^2)s^2
+\frac{7\rt}{20}a^2s^3 \nonumber\\
&f_3=&(\rt+a-\frac{\rt}{3}(1-\rt a -\frac{2}{5}a^2)
+\frac{7\rt}{20}a^2)s^2 \nonumber\\
&f_4=&1/s+\rt as+\frac{1}{3}(1-\rt a -\frac{2}{5}a^2)s^2
-\frac{2}{5}a^2s^3 \nonumber\\
&f_5=&-1/s-\rt as+\frac{1}{3}(1-\rt a -\frac{2}{5}a^2)s^2
+\frac{2}{5}a^2s^3.
\label{asoln}
\eea

The next item to consider is the expression for the
\lq constants\rq, $\alpha,\beta$ and $\gamma$. Of course
they are only constant for exact solutions of the
equations so for our approximate solutions (\ref{asoln}) 
they will have
higher order corrections in $s$. Explicitly
we find that $\alpha=\alpha_0+O(s^4)$, $\beta=\beta_0
+O(s^2)$ and $\gamma=\gamma_0+O(s^2)$ where
\bea
&\alpha_0=&12\rt a \nonumber\\
&\beta_0=&-\frac{1}{3}(4\rt +10a+\frac{29\rt}{10}a^2) 
\nonumber\\
&\gamma_0=&\frac{8}{3}(5-5\rt a -2a^2).
\label{aconsts}
\eea
The second piece of
approximate twistor data is thus the family of 
approximate spectral
curves
\be
\eta^3+\alpha_0\eta\zeta^2+\gamma_0\zeta^3
+i\beta_0(\zeta^6+1)=0.\ee
It is useful to analyse these curves, as it gives an 
indication as to the accuracy of the approximate construction.

The first comparison that can be made is with the
the exact tetrahedral 
3-monopole spectral curve (\ref{tetcurve}). This has 
\be
\alpha=0,\g \gamma=-5\rt\beta, \g
-\beta=\frac{\Gamma(1/6)^3\Gamma(1/3)^3
}{\pi^{3/2}3^38\sqrt{2}}\approx 1.95.
\ee
For the approximate curve to have 
 $\alpha_0=0$ requires $a=0$, then by the above
formulae we obtain
\be
\alpha_0=0,\g \gamma_0=-5\rt\beta_0, \g
-\beta_0=4\rt/3\approx 1.89.
\ee
Note the surprising result that the relation between
$\beta_0$ and $\gamma_0$ is exact in this case, so
that the approximate spectral curve has tetrahedral
symmetry.
The value of $\beta_0$ is also very close
to the true $\beta$ value, considering the 
low order approximation
we chose. Clearly the accuracy of the approximate curves
could be improved by performing an identical construction
as above but taking a higher order approximating series.

By considering asymptotic spectral curves we shall
now determine the range of the parameter $a$, and also
make some further comparisons. The asymptotic monopole
configurations along the exact geodesic can be determined
from the corresponding rational map \cite{HMM}.
At one end the configuration is that of three well-separated
unit charge monopoles on the vertices of a 
large equilateral triangle
in the $x_1x_2$-plane. At the other end the configuration
is asymptotic to  an axisymmetric 2-monopole on the 
negative $x_3$-axis, with the
$x_3$-axis the axis of symmetry, and a unit charge 
monopole on the
positive $x_3$-axis. These two clusters are well-separated
with the distance from the origin of the 1-monopole
being twice that of the 2-monopole.

The spectral curve of a 1-monopole with position 
$(x_1,x_2,x_3)$ is 
\be
\eta-(x_1+ix_2)+2x_3\zeta+(x_1-ix_2)\zeta^2=0
\ee
and is called a star.
By taking a product of such stars the asymptotic
spectral curves corresponding to the above configurations
can be determined \cite{HMM}.
Taking three unit charge monopoles in the $x_1x_2$-plane with
coordinates $x_1+ix_2=ib\omega^j$, $j=0,1,2$ and 
$\omega=e^{2\pi i/3}$
gives the product of stars
\be
0=(\eta-ib(1+\zeta^2))(\eta-ib\omega(1+\omega\zeta^2))
(\eta-ib\omega(\omega+\zeta^2))
=\eta^3+\eta\zeta^23b^2-i(1+\zeta^6)b^3.
\ee
Hence the asymptotic curve has dihedral $D_3$ symmetry,
since $\gamma=0$, with the other two parameters, 
$\alpha$ positive, $\beta$ negative,
satisfying the  relation
\be
\alpha(-\beta)^{-2/3}=3.
\label{xasya}
\ee
We now define the upper limit $a_+$ of the parameter $a$
to be that for which the approximate curve has $D_3$,
symmetry ie. $\gamma_0=0$. By equation 
(\ref{aconsts}) this determines $a_+$ to be
\be
a_+=\frac{3\sqrt{5}-5}{2\sqrt{2}}\approx 0.60.
\ee
At this value of $a$ it can be calculated that
\be
\alpha_0(-\beta_0)^{-2/3}\approx 3.8
\ee
which should be compared with (\ref{xasya}).

The product of stars of a 1-monopole at position
$(0,0,b)$ and an axisymmetric 2-monopole at
$(0,0,-b/2)$ is
\be
0=(\eta+2b\zeta)(\eta^2-2b\eta\zeta+(b^2+\frac{\pi^2}{4})\zeta^2)
=\eta^3+\eta\zeta^2(\frac{\pi^2}{4}-3b^2)+2b(b^2+
\frac{\pi^2}{4})\zeta^3.
\ee
Hence the asymptotic curve is axisymmetric about the
$x_3$-axis, since $\beta=0$. For large separation
the asymptotic relation is that $\alpha$ is negative
and $\gamma$ is positive with
\be
\alpha\gamma^{-2/3}=-\frac{3}{4^{1/3}}\approx -1.9.
\label{xasyb}
\ee
In a similar fashion to above, the lower limit $a_-$
 is defined to be the value of $a$ for which the
 approximate curve has axial symmetry ie. $\beta_0=0$.
This gives
\be
a_-=\frac{\sqrt{2}}{29}(3\sqrt{5}-25)\approx -0.89
\ee
and at this parameter value
\be 
\alpha_0\gamma_0^{-2/3}=\frac{2(87)^{1/3}}{(15
(65-2\sqrt{5}))^{2/3}}\approx -1.7
\ee
which is to be compared with (\ref{xasyb}).

In Figure 1, we plot the approximate spectral curve
coefficients, $\alpha_0$,$\beta_0$,$\gamma_0$ for
$a\in[a_-,a_+]$. 

Having discussed the approximate spectral curves it is
now time to use the approximate Nahm data for its
intended purpose. We take it as input for the numerical
ADHMN construction developed in an earlier paper \cite{HSa}.
Figure 2 shows the output, in the form of a surface
of constant energy density, for each value of the input
 parameter $a$. The surfaces shown correspond to the
values $a=0.4,0.2,0.1,0.0,-0.1,-0.2,-0.4.$
The low energy 3-monopole scattering process is as follows.
Initially, Fig 2.1, there are three unit charge monopoles on
the vertices of a contracting equilateral triangle
in the $x_1x_2$-plane. As they merge each monopole raises
an arm, so that they link in a kind of maypole dance, Fig 2.2.
The legs of the monopoles continue towards the centre,
Fig 2.3, until they too merge and the tetrahedral 3-monopole
is formed, Fig 2.4. Next the top segment of the tetrahedron
separates from the bottom, Fig 2.5, and as it moves up the
$x_3$-axis it leaves behind a torus with three prongs, Fig 2.6.
As the single monopole continues its journey up the
$x_3$-axis it becomes more spherical, and the 2-monopole
smooths out into a torus as it moves down the $x_3$ axis.

\section{Higgs zeros}
\news
Numerical evidence suggests \cite{HSc,S} that the tetrahedral
3-monopole has five zeros of the Higgs field, despite
the fact that it is a charge three monopole.
A scattering geodesic through the tetrahedral 3-monopole
has been investigated in detail and the associated
dynamics, creation and annihilation of the Higgs zeros
tracked numerically \cite{HSc}. 
Since the cyclic scattering of three monopoles 
passes through the tetrahedral 3-monopole, we have an
opportunity to study further this novel phenomenon of
extra Higgs zeros. We compute an analytical approximation
to the Higgs field, and find that it is consistent with
a conjectured behaviour of the Higgs zeros.

Finding the Nahm data effectively solves the nonlinear part 
of the monopole construction but
in order to calculate the Higgs field
the linear part of the ADHMN construction must also be
 implemented. Given
Nahm data $(T_1,T_2,T_3)$ for a $k$-monopole we must solve the 
ordinary differential equation  
\be
({\identity}_{2k}\frac{d}{ds}+{\identity}_k\otimes x_j\sigma_j
+iT_j\otimes\sigma_j){\bf v}=0
\label{lin}
\ee
for the complex $2k$-vector ${\bf v}(s)$,
 where $\identity_k$ denotes
the $k\times k$ identity matrix, $\sigma_j$ are the
 Pauli matrices and
${\bf x}=(x_1,x_2,x_3)$ is the point in space at
 which the Higgs
field is to be calculated. Introducing the inner product
\be
\langle{\bf v}_1,{\bf v}_2\rangle =\int_0^2 {\bf v}_1^\dagger{\bf v}_2\ ds
\label{ip}
\ee
then the solutions of (\ref{lin}) which we require are
 those which are
normalizable with respect to (\ref{ip}). It can be shown
 that the
space of normalizable solutions to (\ref{lin}) has 
(complex) dimension
2. If $\widehat {\bf v}_1,\widehat {\bf v}_2$ 
is an orthonormal basis
for this space then the Higgs field $\Phi$ is given by
\be
\Phi=i\left[ \begin{array}{cc}
\langle(s-1)\widehat {\bf v}_1,\widehat {\bf v}_1\rangle &
\langle(s-1)\widehat {\bf v}_1,\widehat {\bf v}_2\rangle \\
\langle(s-1)\widehat {\bf v}_2,\widehat {\bf v}_1\rangle &
\langle(s-1)\widehat {\bf v}_2,\widehat {\bf v}_2\rangle 
\end{array}
\right].
\label{higgs}
\ee
The strategy adopted is to use the approximate Nahm data
of the previous section and compute approximate solutions
of (\ref{lin}). 

To begin we consider the initial value problem of
(\ref{lin}) at the pole $s=0$, which has the form
\be
s\frac{d{\bf v}}{ds}=B_s{\bf v}
\label{pbc}
\ee
where $B_s$ is a regular $6\times 6$ matrix function of 
$s\in[0,2)$. This is a regular-singular problem and
the eigenvalues of $B_0$ are $\{1,1,1,1,-2,-2\}$, hence
there is a four-parameter family of solutions which are regular
for $s\in[0,2)$. Expressing the solution as a series
\be
{\bf v}(s)=\sum_{j=1}^\infty {\bf d}^{(j)}s^j
\label{serlhs}
\ee
the four arbitrary parameters may be taken to be
 $d_1,..,d_4$ where
\be {\bf d}^{(1)}=
(d_1,d_2,d_3,d_4,i\sqrt{2}d_2,-i\sqrt{2}d_3)^t.\ee
Similarly for the intial value problem at $s=2$ 
there exists a four-parameter family of solutions
$\widetilde{\bf v}(s)$ which are regular for $s\in(0,2]$
and which may be expressed as a series in $\widetilde s=2-s$
\be
\widetilde{\bf v}(s)=
\sum_{j=1}^\infty {\bf h}^{(j)}\widetilde s^j.
\label{serrhs}
\ee
In this case the four arbitrary parameters are contained in
\be
{\bf h}^{(1)}=
(h_1,h_2,h_3,i\sqrt{2}h_1,h_4,ih_3/\sqrt{2})^t.\ee
Taking the solution to be ${\bf v}(s)$ for $s\in[0,1]$
and $\widetilde{\bf v}(s)$ for $s\in[1,2]$ results
in the matching condition
 ${\bf v}(1)=\widetilde{\bf v}(1)$ at $s=1$.
Since this is a vector equation with six components
this gives six constraints involving the eight arbitrary
parameters. We thus obtain the required two parameter 
family of solutions from which to construct an orthonormal
basis. 

To perform the approximate ADHMN construction we
use the approximate Nahm data (with parameter $a$)
 of the previous section
and truncate the series (\ref{serlhs}) and (\ref{serrhs})
at quartic order. Since all functions are now simple
series the required integrals are elementary. 
The above procedure was implemented
using MAPLE. Despite the low order approximations
used the calculations are very involved, and to
simplify matters we restrict to calculating the Higgs
field on the $x_3$-axis only. That is, we set $x_1=x_2=0$
and $x_3=L$ in the above.
In this case a convenient choice of gauge (equivalent
to a choice of orthonormal basis) exists in which
the Higgs field is diagonal
\be
\Phi=i\left[\begin{array}{cc}
\varphi&0\\
0&-\varphi
\end{array}\right].
\ee
The output of the MAPLE program then consists of the
function $\varphi$, which depends on the coordinate
along the $x_3$-axis $L$ and the input parameter $a$,
which determines which monopole configuration along the
geodesic we are studying. Even in this simplified case the 
result is rather complicated
 (the expression fills a full page!).
We plot, in Figure 3, $\varphi$ as a function of $L$ for four
values of $a$. The first (Figure 3.1) is for $a=0$, which is the
tetrahedral 3-monopole. It can be clearly seen that
there are two distinct zeros of the Higgs field along this
line. One is at the origin and the other is at $L\approx 1$,
which is associated with one of the vertices of the
tetrahedron. This result is in good agreement with the
numerical results of \cite{HSc,S}. It confirms that
the tetrahedron has four zeros, each with a local winding
$+1$, on the vertices of a tetrahedron and a fifth zero,
which we refer to as an anti-zero since it has local
winding $-1$, at the origin. 

The remaining plots in Figure 3 are evidence to support
the following conjecture on the dynamics of the Higgs
zeros. Initially the three monopoles are well-separated,
so there can only be three zeros. They are positioned
on the vertices of an equilateral triangle in the 
$x_1x_2$-plane. 
At all times the zeros of the Higgs field must be
consistent with the $C_3$ symmetry, which leads
us to the following description.

The three zeros on the vertices of a triangle approach
the $x_3$-axis, but fall below the origin.
During their approach there is a critical point
on the $x_3$-axis where
there is a zero anti-zero creation event, sending
an anti-zero down the $x_3$-axis and a plus zero up
the $x_3$-axis (Figure 3.2, $a=0.05$).
The anti-zero reaches the origin (this
is the tetrahedral 3-monopole, Figure 3.1, $a=0$) 
and continues down
the $x_3$-axis (Figure 3.3, $a=-0.05$) until it 
eventually meets up with
the three positive zeros as they hit the $x_3$-axis.
Then we have lost all anti-zeros and are left with
the $+1$ zero
of the unit charge monopole moving up the $x_3$-axis
and the $+2$ zero of the axisymmetric 2-monopole moving
down the $x_3$-axis. As a last check we verify that
for $a$ sufficiently large there are no zeros on the
$x_3$-axis (Figure 3.4, $a=0.2$).

From Figure 3 we see that the number of zeros on the
$x_3$-axis has a simple interpretation. The shape of
the $\varphi$ curve remains the same, but as $a$ decreases
the curve moves up the plane. Thus the number of 
roots of this curve increases from zero to 
an instantaneous double root and
finally to two simple roots, 
with the separation between the two roots
then monotonically increasing.

We now address the question of the uniqueness of the
above interpretation of the motion of the Higgs
zeros drawn from the data. It turns out that there
is a second possibility which is more complicated than
the first and involves a splitting into anti-zeros
of each of the three initial Higgs zeros. The results
given so far are consistent with both conjectures
and an additional check must be made to decide between
the two. For the first possibility, which will turn
out to be the correct one, we have seen that 
for configurations just prior to the formation
of the tetrahedral 3-monopole there are two zeros
on the $x_3$-axis which have a local winding number $+1$
and $-1$ respectively. An analysis of the second possibility
demands that these zeros have local winding numbers
$+1$ and $+2$. To settle the issue a numerical calculation
is performed. First we select a suitable configuration,
which is taken to be $a=0.02$. Figure 4 shows a plot of the 
length squared of the Higgs field (solid line) and a component
of the Higgs field (dashed line) along the axis $x_3=L$.
From this the positions of the two zeros can be
approximately read off as $L\approx 0.6$ and $L\approx 0.15$.
Next we apply the numerical algorithm given in 
\cite{HSc} to compute the local winding number
$Q_R(L)$ of the normalized Higgs field on a two-sphere
of radius $R$, with centre $(x_1,x_2,x_3)=(0,0,L)$.
As a first check we compute that $Q_2(0)=+3$, so that
when all zeros are inside the chosen two-sphere the
local winding number indeed gives the number of monopoles.
Next we examine the zero which is highest on the $x_3$-axis
and confirm that $Q_{0.1}(0.6)=+1$. Finally, the 
computation that $Q_{0.1}(0.15)=-1$ shows that the
second zero on the $x_3$-axis is an anti-zero and
thus rules out the second possibility for the
motion of the Higgs zeros to which we alluded above.

For geodesics involving monopoles which are linearized
on a quotient curve which is elliptic, it has been
conjectured \cite{HSc,S} that the special 
\lq splitting points\rq, where anti-zeros appear/disappear,
are associated to singular points where an elliptic
curve becomes rational. Presumably when the geodesic
involves monopoles which are linearized on a higher
genus surface (as here) the \lq splitting points\rq\ 
will correspond to vanishing cycle points of the
surface.

\section{Cyclic 4-monopoles}
\news
For $C_4$ symmetric 4-monopoles there are three
types of geodesics, corresponding to the surfaces
$\Sigma_4^0,\Sigma_4^1$ and $\Sigma_4^2$. The first of
these is the essentially planar scattering where the
four monopoles instantaneously form an axially
symmetric configuration and subsequently
scatter through an angle of $45^\circ$. We shall not
be concerned with this type of scattering. The other two
processes are fully three-dimensional and are of more interest.
The $\Sigma_4^2$ geodesic has an increased symmetry
from $C_4$ to $D_4$ and describes a scattering through
the cubic 4-monopole \cite{HMM}. It is possible to 
investigate this geodesic in exactly the same manner
as was done earlier for the $C_3$ geodesic through the 
tetrahedral 3-monopole. However, the $\Sigma_4^1$ geodesic does
not contain any Platonic monopole configurations, and so at
first sight it appears that we require a new ingredient
to deal with this case; since we do not know the exact
Nahm data at any point on the geodesic we can not read-off
the residue behaviour. It turns out that it is instructive
not to deal with each case separately, but to
apply a unified treatment. This is presented in the following.

First of all, we use equation (\ref{toda}) to obtain
the form of the Nahm data from the algebra $A_{3}^{(1)}$.
It is convenient to change notation and write
$$
T_1=\frac{1}{2}\left[\begin{array}{cccc}
0&-q_1&0&q_0\\
q_1&0&-q_2&0\\
0&q_2&0&-q_3\\
-q_0&0&q_3&0\end{array}\right]; \
T_2=-\frac{i}{2}\left[\begin{array}{cccc}
0&q_1&0&q_0\\
q_1&0&q_2&0\\
0&q_2&0&q_3\\
q_0&0&q_3&0\end{array}\right]; \
$$
\be
T_3=i\left[\begin{array}{cccc}
p_1&0&0&0\\
0&p_2&0&0\\
0&0&p_3&0\\
0&0&0&p_0
\end{array}\right].
\label{data4}
\ee
Nahm's equation then becomes the set of equations
(equivalent to the 4-particle Toda chain)
\be
\dot p_j=\frac{1}{2}(q_j^2-q_{j-1}^2), \hskip 10pt
\dot q_j=q_j(p_j-p_{j+1})
\label{toda4}
\ee
where the indices are to be read modulo 4, in accordance
with the periodicity of the chain. Since we are considering
strongly centred monopoles we also impose the standard
Toda chain zero momentum relation
\be
\sum_{j=0}^3 p_j=0.
\ee
We shall generally use this relation to eliminate $p_0$
from the equations, but for certain purposes it is
useful to preserve the symmetry of the system and not
explicitly solve for $p_0$.

The spectral curve of the Nahm data (\ref{data4})
is 
\be
\eta^4+\alpha\eta^2\zeta^2+\beta(\zeta^8+1)+\gamma\zeta^4+\delta
\eta\zeta^3=0
\ee
where the real constants are
\bea
\alpha&=&\sum_{j=0}^3\{q_j^2+2p_j(2p_{j+1}+p_{j+2})\}\\
\beta&=&-\prod_{j=0}^3 q_j\\
\gamma&=&16\prod_{j=0}^3 p_j+\frac{1}{2}
\sum_{j=0}^3\{q_j^2(q_{j+2}^2+8p_{j-1}p_{j+2})\}\\
\delta&=&\sum_{j=0}^3\{q_j^2(p_{j-1}+p_{j-2})
+8p_jp_{j+1}p_{j-1}\}.
\eea

We now examine the residue behaviour at $s=0$.
Putting
\be
q_j\sim r_j/s, \hskip 10pt p_j\sim R_j/s 
\g \mbox{as} \ s\rightarrow 0
\ee
into equation (\ref{toda4}) gives
\bea
2R_j+r_j^2-r_{j-1}^2&=&0\label{res1}\\
r_j(R_j-R_{j+1}+1)&=&0.\label{res2}
\eea
Substituting the first of these equations into the second
we obtain
\be
r_j(r_{j+1}^2-2r_j^2+r_{j-1}^2+2)=0.
\ee
Now if all the $r_j$ are non-zero then, recognising
the bracket as the second order difference approximation
to the second derivative, the general solution of the
difference equation is
\be
r_j^2=c_0+c_1j-2j^2
\ee
for constants $c_0,c_1$. However, this solution does
not satisfy the periodic boundary conditions $r_{j+4}=r_j$,
and thus we have proved that at least one of the $r_j$'s
must be zero. 

By symmetry we can choose $r_0=0$. All the $R_j$'s must
be non-zero (in order for the matrix residues to
form the irreducible 4-dimensional representation of $su(2)$)
which together with equation (\ref{res1}) implies that
$r_1r_3\ne 0$. Furthermore, $r_2\ne 0$, since if $r_2=0$
then the solution is $r_1^2=r_2^2=1$ which by
equation (\ref{res1}) determines
$(R_0,R_1,R_2,R_3)=(1/2,-1/2,1/2,-1/2)$ giving the
reducible representation $\underline{2}\oplus\underline{2}$.
Now that we have $r_0=0$ and $r_1r_2r_3\ne 0$ the equations
simplify to the linear system
\be
\left[\begin{array}{ccc}
2&-1&0\\
-1&2&-1\\
0&-1&2\end{array}\right]
\left[\begin{array}{c}
r_1^2\\
r_2^2\\
r_3^2
\end{array}\right]
=\left[\begin{array}{c}
2\\
2\\
2
\end{array}\right].
\ee
Note that the matrix occuring in the above is the
Cartan matrix of the Lie algebra $A_3$. The unique
solution is
\be
r_0=0, \g r_1^2=3, \g r_2^2=4, \g r_3^2=3
\label{res1soln}
\ee
which gives from equation (\ref{res1}) that
\be
(R_0,R_1,R_2,R_3)=(3/2,-3/2,-1/2,1/2)
\label{res2soln}
\ee
which identifies the matrix residues as forming the
required irreducible representation $\underline{4}$.

The upshot is that the only freedom in the choice
of residues is which one of the $r_j$'s is chosen
to be zero. There are some choices of sign available
in taking the square roots in (\ref{res1soln}) but
these only lead to an inversion of the whole configuration
and are of no importance. By a suitable choice of gauge
it is always possible to choose $r_0=0$ 
and so at first glance it may appear a puzzle as to how the 
different types of geodesic are selected. However, there
is a similar second pole at $s=2$ at which
again one of the associated residues $\widetilde r_j$
 must be zero. Since we
are not using a basis in which the symmetry property
$T_i(2-s)=T_i(s)^t$ of the Nahm data is manifest, then once
the gauge freedom has been used to fix $r_0=0$ then we have
 no freedom left to specify which
$\widetilde r_j$ vanishes. The different geodesics
are distinguished by having different $\widetilde r_j$'s
 being zero. Another way to view this is that the
explicit symmetry transformation relating Nahm data at $s$ to
Nahm data at $2-s$ will be different for different geodesics
in the same basis. Of course, by general arguments it
is true that a basis exists in which the symmetry property
is the manifest one given above, but the important point
is that this basis is not the same one for different geodesics.
This means that there is no point looking for
a general \lq\lq good basis\rq\rq\ at the start of the
 calculation, since this will  vary with each solution.
It does however mean that all the geodesics can be treated
at once and the different cases identified near the 
end of the calculation by the imposition of the different
symmetry properties.

We now construct the approximate twistor data, using the
method introduced earlier. A particular solution, which is
a single-valued expansion about the $s=0$ pole with
residues $r_j,R_j$ as determined above, is constructed for the functions
$q_j,p_j$. We omit the details since the method is just
as in the $C_3$ case considered earlier. Applying
Painlev\a'e analysis gives a linear system with the matrix
determinant 
\be
\Delta=8(r+1)(r+2)(r+3)(r-2)(r-3)(r-4)^2
\ee
giving the Kowalevski exponents. The solution thus
contains four arbitrary constants, which is the required
number to coincide with the constants
$\alpha,\beta,\gamma,\delta$ in the spectral curve.

Again we work with expansions to cubic order in $s$
and take these as the approximate Nahm data for $s\in[0,1]$.
To determine a one-parameter family of
monopoles we must impose a symmetry relation to
construct the Nahm data for the second half of the interval ie.
$s\in[1,2]$. \\

{\underline{\sl The case $\Sigma_4^2$}}\\

For this geodesic the symmetry property of equations
(\ref{toda4}) which we exploit is the transformation
under $s\mapsto 2-s$ of
\bea
(q_0,q_1,q_2,q_3)&\mapsto&(q_2,q_1,q_0,q_3)\\
(p_0,p_1,p_2,p_3)&\mapsto&(p_3,p_2,p_1,p_0).
\eea
Using this transformed data in the second half of
the $s$ interval gives the matching conditions at $s=1$
\be
q_0(1)=q_2(1), \g p_1(1)=p_2(1), \g 
p_1(1)+p_2(1)+2p_3(1)=0.
\ee
Applying these three constraints to the four-parameter
family of solutions we arrive at one of the sought after 
one-parameter families. The constant terms in the
spectral curve coefficients are given by
\bea
\alpha_0&=&20\sqrt{3}a\\
\beta_0&=&\frac{15}{2}+7\sqrt{3}a+4a^2\\
\gamma_0&=&105-70\sqrt{3}a+80a^2\\
\delta_0&=&0.
\eea
In fact it is easy to see that the imposed symmetry
forces the reductions $q_1=q_3$ and $p_3=-p_2$ which
in turn gives $\delta=0$ (not just $\delta_0=0$).
 Thus the approximate data has exact $D_4$ symmetry.
We therefore already know that we are considering either the
$\Sigma_4^0$ or the $\Sigma_4^2$ geodesic. The confirmation
that we have the second of these two possibilities is
provided by examination of the approximate spectral curve
for $a=0$. This curve is given by
\be
\alpha_0=0, \g \gamma_0=14\beta_0, \g \beta_0=15/2.
\ee
The first two of these relations are enough to
ensure that this curve has cubic symmetry. Thus we
have a geodesic containing a cubic configuration
and hence are considering the $\Sigma_4^2$ geodesic.
For comparison the exact cubic 4-monopole has a  curve
given by\footnote{There is a factor of 16 error
 in ref. \cite{HMM}}
\be
\alpha=0, \g \gamma=14\beta, \g \beta=3\Gamma(1/4)^8/(1024\pi^2).
\ee
Because of the way that $\beta$ enters into the
spectral curve it is really the fourth root 
of $\beta$ which determines the length scale of the 
cubic monopole. Thus the appropriate comparison is
\be
\beta^{1/4}=1.73, \g \beta_0^{1/4}=1.65
\ee
which again demonstrates the reasonable accuracy of such 
a low order approximation scheme.

Using the same kind of analysis as in the $C_3$ case considered
 earlier, it is possible to determine the range of the 
parameter $a$ by examining the limiting spectral curves.
We apply the numerical ADHMN construction to this
approximate Nahm data for the values $a=0.4,0.2,0.1,0.0,-0.1,
-0.2,-0.4$. The resulting energy density surfaces are shown in
Figure 5. The four monopoles approach on the vertices
of a contracting square Fig 5.1, and bend as they merge
Fig 5.2, until they form the cube Fig 5.4. The top and
bottom portions of the cube then pull apart Fig 5.5, and
become increasingly toroidal as the pair of charge 2
monopoles separate along the $x_3$-axis Fig 5.7.\\

{\underline{\sl The case $\Sigma_4^1$}}\\

For this geodesic the relevant symmetry property of
equation 
(\ref{toda4}) is the transformation
under $s\mapsto 2-s$ 
\bea
(q_0,q_1,q_2,q_3)&\mapsto&(q_1,q_0,q_3,q_2)\\
(p_0,p_1,p_2,p_3)&\mapsto&(p_2,p_1,p_0,p_3)
\eea
leading to the three matching conditions
\be
q_0(1)=q_1(1), \g q_2(1)=q_3(1), \g 
p_1(1)+2p_2(1)+p_3(1)=0.
\ee

In this case the approximate spectral curve coefficients
are a little more complicated
\bea
\alpha_0&=&20\sqrt{3}a\\
\beta_0&=&-\frac{30\sqrt{3}+a(138+72\sqrt{3})
+a^2(36+25\sqrt{3})}{9+5\sqrt{3}}\\
\gamma_0&=&\frac{
-1260-1680\sqrt{3}+a(3360+2100\sqrt{3})
+a^2(4530+2500\sqrt{3})}{(\sqrt{3}+2)(9+5\sqrt{3})}\\
\delta_0&=&\frac{
-2400-1200\sqrt{3}+a(2400+1360\sqrt{3})+a^2(520+320\sqrt{3})}
{(\sqrt{3}+2)(9+5\sqrt{3})}.
\eea
The salient feature is that $\delta_0\ne 0$, so the
symmetry is $C_4$ and not $D_4$, thereby identifying
the geodesic as the one associated with the surface
$\Sigma_4^1$.

In figure 6 we show energy density plots for the
parameter values 

$a=0.6,0.5,0.4,0.3,0.2,0.0$.
This scattering appears very similar to the $C_3$
scattering discussed earlier. The four monopoles
approach on the vertices of a contracting square
Fig 6.1, and as they merge they link arms forming a
pyramid shaped configuration Fig 6.3. The top of
the pyramid breaks off, Fig 6.5, and travels up the $x_3$-axis,
while the charge three base travels down the $x_3$-axis. 
As they continue to separate, Fig 6.6, the unit charge monopole
 becomes more spherical and the base deforms closer towards the
axisymmetric 3-monopole.\\

Numerical evidence suggests \cite{S} that the cubic
4-monopole does not possess anti-zeros, in contrast
 to the tetrahedral 3-monopole. Figures 5 and 6 appear
to add some understanding to this result, since it is
the $\Sigma_4^1$ geodesic, rather than the $\Sigma_4^2$
geodesic,
which appears to be the closest 4-monopole analogue
of the $\Sigma_3^1$ geodesic. Hence we might expect
the pyramid-like 4-monopole, rather than the cubic 4-monopole
to be the one which has anti-zeros.

\section{Conclusion}
\news

We have proposed an analytical method to obtain
approximate Nahm data which, when combined with the previously
introduced numerical ADHMN construction, provides an efficient 
approximation scheme for
the construction of monopoles. This has been applied to the study
of monopoles with cyclic symmetry and shown to produce good results.
The scattering processes reveal exotic dynamics, and
indicate that the key to understanding this type of scattering lies
with a more detailed knowledge of the zeros of the Higgs field.\\

\noindent{\bf Acknowledgements}

Many thanks to Nigel Hitchin, Conor Houghton, Nick Manton and
Andrew Pickering for useful discussions.

\newpage
\noindent{\bf Figure Captions}\\

{\bf Fig 1.}
Approximate spectral curve coefficients\\

{\bf Fig 2.}
$C_3$ symmetric energy density surfaces.
The parameter values are \g\g
$a=0.4,0.2,0.1,0.0,-0.1,-0.2,-0.4$\\

{\bf Fig 3.}
Plot of the Higgs component $\varphi$
on the $x_3$-axis, for configurations with
$(1) a=0;\ (2) a=0.05; \ (3) a=-0.05; \ (4) a=0.2.$\\

{\bf Fig 4.}
Plots of the length squared of the Higgs field
(solid line) and a Higgs component (dashed line) along
the $x_3$-axis for $a=0.02$\\

{\bf Fig 5.}
$D_4$ symmetric energy density surfaces.
The parameter values are \g $a=0.4,0.2,0.1,0.0,-0.1,-0.2,-0.4$\\

{\bf Fig 6.}
$C_4$ symmetric energy density surfaces.
The parameter values are \g
$a=0.6,0.5,0.4,0.3,0.2,0.0$\\

Note: Figures 2,5 \& 6 are attached gif files
cyclicfig2.gif etc.

\newpage
\begin{figure}[ht]
\begin{center}
\leavevmode
{\epsfxsize=15cm \epsffile{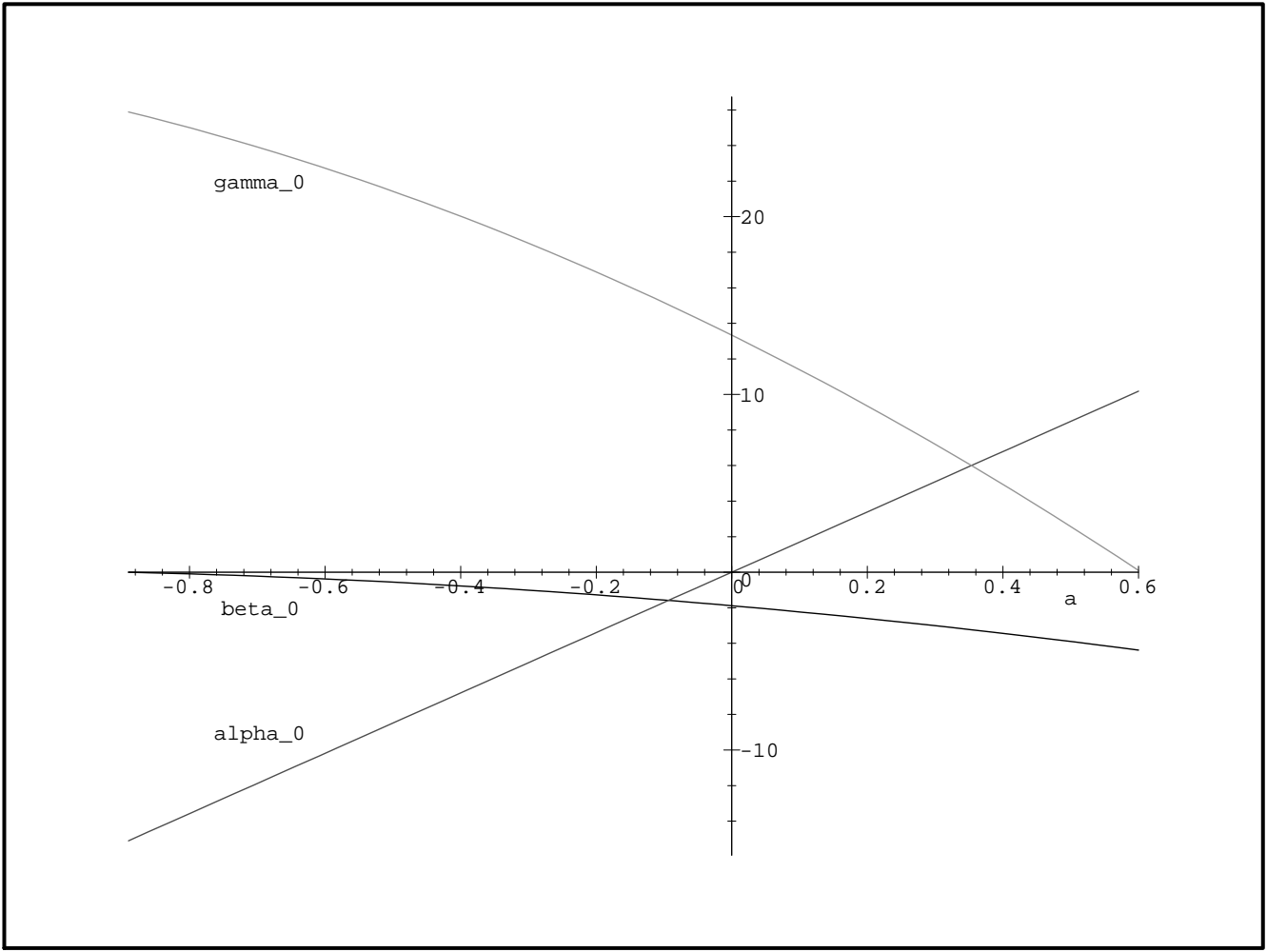}}
\end{center}
\caption{Approximate spectral curve coefficients}
\end{figure}

\begin{figure}[ht]
\caption{View figure by typing {\sl xv cyclicfig2.gif}}
\end{figure}

\newpage
\begin{figure}[ht]
\begin{center}
\leavevmode
{\epsfxsize=15cm \epsffile{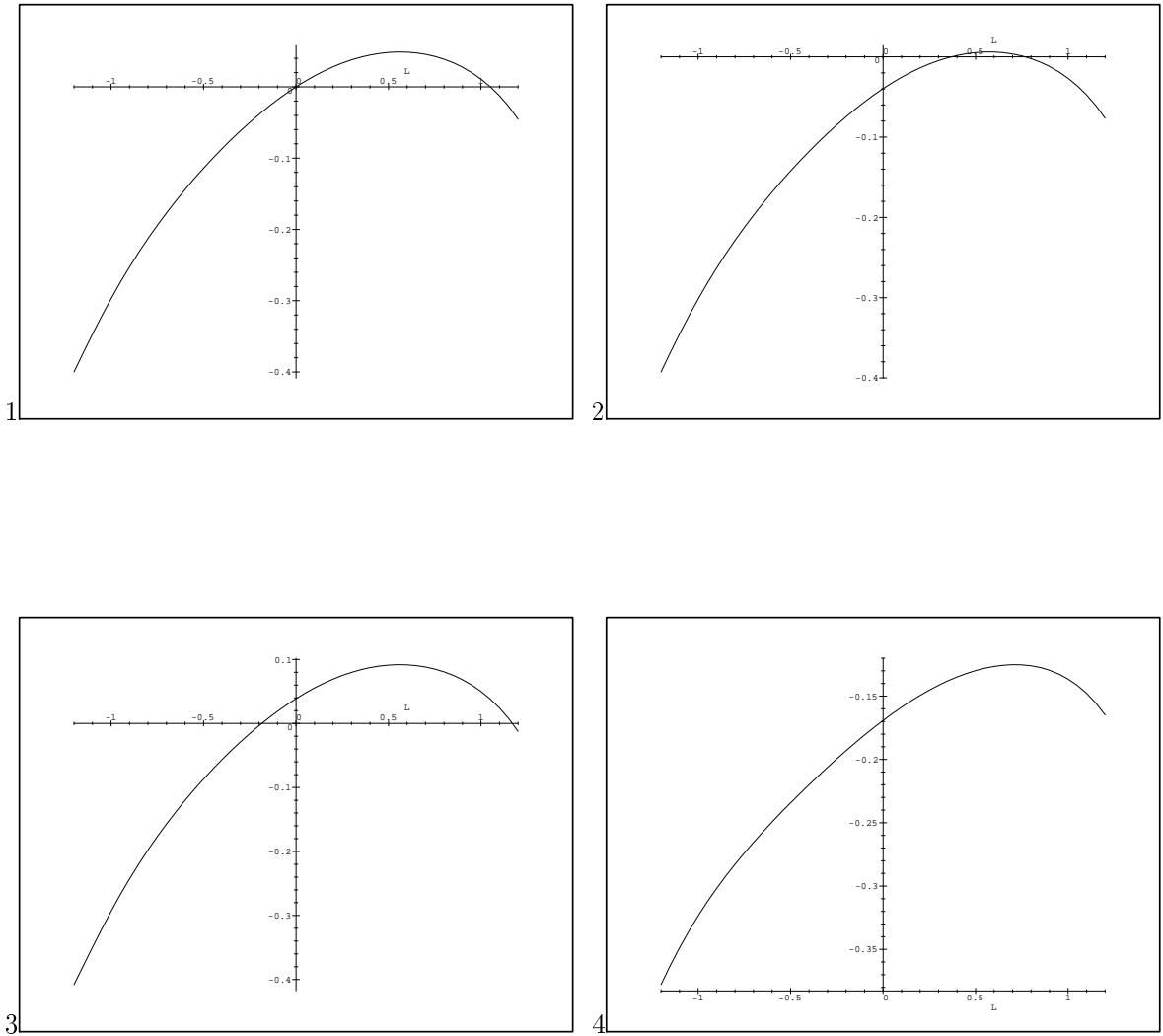}}
\end{center}
\ \vskip -2cm
\caption{Plot of the Higgs component $\varphi$
on the $x_3$-axis, for configurations with
$(1) a=0;\ (2) a=0.05; \ (3) a=-0.05; \ (4) a=0.2.$}
\end{figure}

\newpage
\begin{figure}[ht]
\begin{center}
\leavevmode
{\epsfxsize=12cm \epsffile{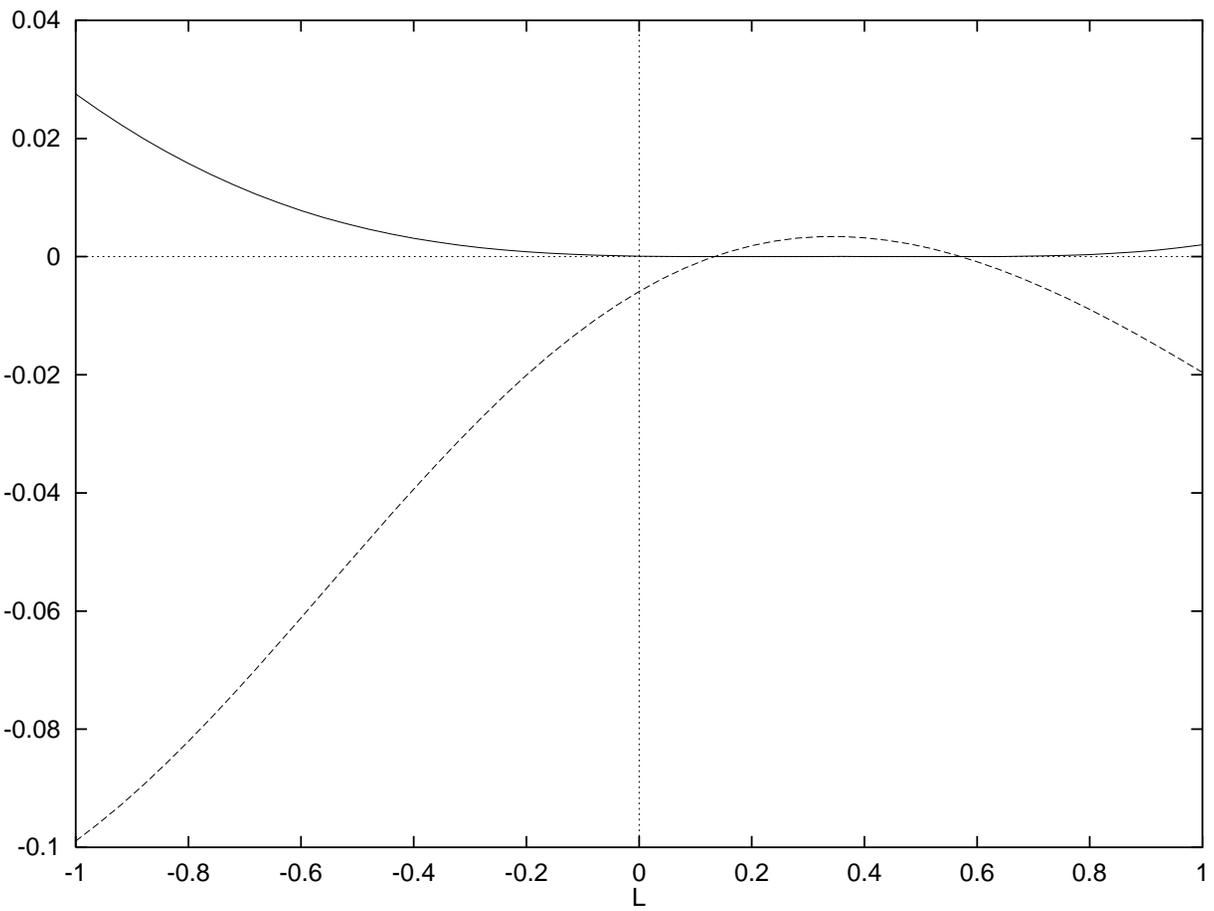}}
\end{center}
\caption{Plots of the length squared of the Higgs field
(solid line) and a Higgs component (dashed line) along
the $x_3$-axis for $a=0.02$}
\end{figure}

\end{document}